\magnification=\magstep1 \overfullrule=0pt 
\advance\hoffset by -0.35truecm   
\vsize=23.3truecm
\font\tenmsb=msbm10       \font\sevenmsb=msbm7
\font\fivemsb=msbm5       \newfam\msbfam
\textfont\msbfam=\tenmsb  \scriptfont\msbfam=\sevenmsb
\scriptscriptfont\msbfam=\fivemsb
\def\Bbb#1{{\fam\msbfam\relax#1}}
\def\Z{{\Bbb Z}}

\font\grosss=cmr8 scaled \magstep4
\font\grossk=cmr7 scaled \magstep3

\font\sf=cmss10   
\font\klein=cmr8  \font\itk=cmti8  \font\bfk=cmbx8      
\def\lb{\lbrack}\def\rb{\rbrack}  \def\q#1{$\lb${\rm #1}$\rb$}
\def\bn{\bigskip\noindent} \def\mn{\medskip\smallskip\noindent}
\def\sn{\smallskip\noindent} 
\def\bzeta{\bar\zeta}
\def\one{{\bf 1}}
\def\h{\hbox{$\cal H$}}   
\def\cedille#1{\setbox0=\hbox{#1}\ifdim\ht0=1ex \accent'30 #1%
 \else{\ooalign{\hidewidth\char'30\hidewidth\crcr\unbox0}}\fi}

\def\hbn{\hfill\break\noindent}
\def\scst{\scriptstyle}  \def\ts{\textstyle}
\def\bda{|\!|a\rangle\!\rangle}\def\bdb{|\!|b\rangle\!\rangle}
\def\bdA{|\!|A\rangle\!\rangle}\def\bdbadj{\langle\!\langle b|\!|}
\def\ik{I_{\rm Kac}}
\def\uplc{^{L_0-{c\over 24}}}
\def\tqlc{{\tilde q\uplc}}
  
\def\Fus#1#2#3#4#5#6{
 {\hbox{\sf F}}_{#1#2} \raise-.5pt\hbox{$\bigl\lb$}
 \raise.5pt\hbox{ ${\scst{ {\hfil\!#3\hfil\;\hfil#4\hfil} \atop
           {\hfil\!#5\hfil\;\hfil#6\hfil} }} $}
  \raise-.5pt\hbox{$\bigr\rb$}   }
%
\global\newcount\glgnum
\global\glgnum=0
\def\glg{{{\global\advance\glgnum by1}{(\number\glgnum)}}}
\def\mkg#1{\glg\xdef#1{(\the\glgnum)}}
\global\newcount\refnum
\global\refnum=0
\def\preref{{\global\advance\refnum by1}{\number\refnum}}
\def\ref#1{\preref\xdef#1{\the\refnum}}
\def\eq{\eqno\mkg} 
\def\mkgho#1{\xdef#1{(\the\glgnum}}
\def\mkgvo#1{\xdef#1{\the\glgnum)}}
\def\mkgvho#1{\xdef#1{\the\glgnum}}
%
%
{\nopagenumbers
\line{AEI-2000-3 \hfill BONN-TH-2000-02}
\line{IASSNS-HEP-00/21 \hfill hep-th/0003110} 
\sn
\bn\bn\bn
\centerline{
{\grosss On relevant boundary perturbations of}} 
\smallskip
\centerline{
{\grosss unitary minimal models}}
\bn\bn\bn
\centerline{
{\grossk A.\ Recknagel}$\;{}^{1}\,$,\ \ {\grossk D.\ Roggenkamp}$\;{}^{2}\,$
\ \ {\bf and} \ \ {\grossk V.\ Schomerus}$\;{}^{3,4}$}
\bn\bn\bn
\centerline{${}^1\;$Max-Planck-Institut f\"ur Gravitationsphysik}
\centerline{Albert-Einstein-Institut}
\centerline{ Am M\"uhlenberg 1, D-14476 Golm, Germany}
\bn
\centerline{${}^2\;$Physikalisches Institut der Universit\"at Bonn}
\centerline{Nu\ss allee 12, D-53115 Bonn, Germany}
\bn
\centerline{${}^{3}\;$II. Institut  f\"ur Theoretische Physik,
Universit\"at Hamburg,}
\centerline{Luruper Chaussee 149, D-22761 Hamburg, Germany}
\bn
\centerline{${}^{4}\;$ Institute for Advanced Study, School of 
Natural Sciences,}
\centerline{Olden Lane, Princeton, NY 08540, U.S.A.}
\bn\bn\vfill\vfill\bn
\centerline{\bf Abstract}
\bigskip{\noindent
{\narrower\narrower  We consider unitary Virasoro minimal models on the 
  disk with Cardy boundary conditions and discuss deformations 
  by certain relevant boundary operators, analogous to tachyon 
  condensation in string theory. Concentrating on the least 
  relevant boundary field, we can perform a perturbative analysis of 
  renormalization group fixed points. We find that the systems always 
  flow towards stable fixed points which admit no further (non-trivial) 
  relevant perturbations. The new conformal boundary conditions are in 
  general given by superpositions of 'pure' Cardy boundary 
  conditions.} 
}

\bn\bn
\vfill\vfill\vfill
\medskip 
\leftline{e-mail addresses: {\tt anderl@aei-potsdam.mpg.de, 
   roggenka@th.physik.uni-bonn.de,}}
\leftline{\phantom{e-mail addresses: }{\tt vschomer@x4u2.desy.de}}
\eject
}
\pageno=1
\noindent
{\bf 1. Introduction}
\mn
Conformal field theory on surfaces with boundaries, or boundary 
CFT for short, has recently attracted new interest  because it 
provides the framework for a world-sheet analysis of D-branes 
in string theory. It has also applications to various systems 
of condensed matter physics such as the three-dimensional 
Kondo effect \q{\ref{\AfLtwo}}, fractional quantum Hall fluids (see 
e.g.\ \q{\ref{\FLuS}}) and other quantum impurity problems. 
\hbn
A lot of progress has been made in clarifying the rather rich 
intrinsic structure of boundary CFTs, but deformations away from the 
conformal point have been investigated less systematically in the 
literature. Their study should provide some insight into the 
structure of the space of boundary theories and its renormalization 
group fixed points, besides possible applications in condensed matter 
physics. Such deformations are also important in the string theory 
context, where flows triggered by a relevant boundary operator 
appear as tachyon condensation and may result e.g.\ in the formation 
of (non-BPS) bound states of branes, see \q{\ref{\GNS},\ref{\Sen}} and 
the more recent string field theory considerations in 
\q{\ref{\SZB},\ref{\HKM}}. 
\hbn
Most of the existing literature on relevant boundary deformations 
rests on the thermodynamic Bethe ansatz (TBA), on scattering matrices 
or on the truncated conformal space approach (TCSA), see e.g.\ 
\q{\ref{\TBAetc},\ref{\Chim}}. These methods can provide 
(non-perturbative) information about RG flows even between the fixed 
points, but they do not lend themselves easily to model independent 
investigations. Some general aspects of 
conformal perturbation theory have been discussed in the work 
of Affleck and Ludwig \q{\AfLtwo,\ref{\AfLone}}, leading to the formulation 
of the so-called ``g-theorem''. Here, we want to use and refine 
these techniques to study specific relevant boundary deformations 
of arbitrary unitary Virasoro minimal models. The problem we attack 
can be viewed as a boundary analogue of Zamolodchikov's analysis 
of relevant bulk perturbations of minimal models \q{\ref{\ZaPert}}
which showed that deformation of the minimal model $M_m$ by the 
least relevant bulk field $\varphi_{(1,3)}$ induces an RG flow to a 
new fixed point corresponding to the model $M_{m-1}$.  
\hbn
In order to identify the CFT at the new fixed point, 
Zamolodchikov simply checks that, after the deformation, the 
central charge agrees with $c_{m-1}$ up to higher order 
corrections in the coupling constant $\lambda$. It is a
(technically important) peculiarity of the least relevant 
field $\varphi_{(1,3)}$ that the expansion in $\lambda$ 
coincides with an expansion in the parameter $1/m$. 
\hbn
In the study of boundary perturbations, the quantity of interest is 
the ground-state degeneracy $g_{a} = \langle\one\rangle_a$, i.e.\ the 
vacuum expectation value of the identity in the presence of the boundary 
condition $a$, which also plays the main role in the g-theorem. 
In close analogy to \q{\ZaPert}, we first determine the new fixed 
point from the beta function -- to low order in the coupling 
constant or, at the same time, to low order in $1/m$. (The same 
``peculiarity'' as in the bulk case can be exploited for the least 
relevant boundary operator $\psi_{(1,3)}$.) Then we calculate the 
change of $g_a$ in an expansion in $1/m$. Since the set of possible 
boundary conditions for Virasoro minimal models is discrete, the first 
few terms suffice to identify the boundary condition which is 
reached at the perturbative fixed point of the $\psi_{(1,3)}$-flow 
-- up to some residual ambiguities which can be resolved by studying 
one-point functions of other primary bulk fields. 
\sn
In the next section, we recall some facts about minimal models 
on the disk, postponing some lengthier formulas to 
the Appendix. The perturbative analysis of relevant boundary 
deformations is carried out in Section 3, while Section 4 contains 
conclusions and remarks on possible generalizations. 
\sn
On a qualitative level, there are two main lessons to be drawn from 
the results of Section 3. First of all, the RG flows generically 
end up with conformal boundary conditions that can only be described 
by {\sl superpositions} of Cardy boundary states. It appears, therefore, 
that one cannot attribute a more fundamental meaning to such ``pure'' 
boundary conditions (with a unique vacuum state in the spectrum) than 
to superpositions thereof. Both types occur on an equal footing in 
the space of conformal boundary conditions associated with a given 
CFT on the Riemann sphere. Similar effects show up in string theory, 
where single D-branes can be continuously deformed into systems of 
several branes, see \q{\ref{\RSmod},\Sen}. The physical meaning of 
these ``systems'' in terms of two-dimensional phase diagrams can, 
however, be very different: In the recent work \q{\ref{\Aff}}, Affleck 
has connected superpositions of boundary conditions with long range 
order of boundary spins, and thus to first order phase transitions. 
While \q{\Aff} focuses on the tri-critical Ising model ($m=4$), 
this interpretation certainly generalizes to the boundary states 
we will meet below, simply because the main ``world-sheet signal'' 
for an ordered phase, the occurrence of additional dimension zero 
operators, is present in all these cases. 
\hbn
The second comment we would like to make is that some naive
expectations about RG flows fail in the context of boundary 
perturbations: Renormalization can be thought of as integrating 
out degrees of freedom, and intuitively one would expect that the 
more relevant the perturbing operator is the more degrees of freedom 
are integrated out. In the bulk case \q{\ZaPert}, the CFT reached 
after perturbing a minimal model with the least relevant field 
does contain further relevant operators, in accordance with this 
simple picture. In contrast, the boundary flows found in the 
present paper, and in fact most examples studied in the literature 
so far, lead to {\sl stable} fixed points (in the sense that no 
relevant boundary operators remain in the spectrum). On the other 
hand, there are counterexamples like in the three-states Potts model 
\q{\ref{\AOS}} and in SU(2) WZW models \q{\ref{\ARS}}. A perhaps even 
more counterintuitive behaviour of boundary perturbations was discovered 
in \q{\RSmod,\Sen}: In some situations, a (seemingly irreversible) 
relevant boundary perturbation can be ``undone'' by a sequence of 
(invertible) marginal deformations in the bulk and on the boundary.
Such ``failures'' of the Wilsonian picture of RG flows are most 
probably due the the very different weights of bulk and boundary 
degrees of freedom, but we feel that the phenomenon deserves further 
study. 
\bn\mn 
{\bf 2. Boundary conditions for minimal models} 
\mn
We consider unitary Virasoro minimal models $M_m$ with central charges 
$c_m = 1 - {6\over m(m+1)}$ for $m=3,4,\ldots$, and with diagonal 
modular invariant partition functions. The left- and right-moving 
conformal dimensions of the primaries $\varphi_{i}(\zeta,\bzeta)$ 
are given by the Kac table 
$$
h_m(r,s) = { ((m+1)r-ms)^2-1 \over 4 m (m+1)} 
\eq\Kact$$
with $i\equiv (r,s) \in I_{\rm Kac} := \{\,(\bar r,\bar s)\,|\, 
1\leq \bar r \leq m-1,\; 1\leq \bar s \leq m\,\}/\sim\,$, 
where ``$\sim$'' denotes the conformal grid symmetry 
$(r,s) \sim (m-r,m+1-s)$. Conformal boundary conditions 
on such a ``bulk CFT'' can be described by boundary states 
$$
\bda = \sum_{i\in \ik} \; B^i_a\,|i\rangle\!\rangle \ \ .
\eq\bdstgen$$
To each irreducible representation of the Virasoro algebra, there 
is an associated Ishibashi state $|i\rangle\!\rangle=|r,s\rangle\!\rangle$
satisfying linear conditions $\bigl(\,L_n - \overline{L}_{-n}\,\bigr)\,
|i\rangle\!\rangle = 0$ which guarantee conformal invariance of 
the system on the upper half-plane or the unit disk, see 
\q{\ref{\Ish},\ref{\CaFus}} for the construction. The complex 
coefficients $B^i_a$ are subject to non-linear equations like 
sewing constraints \q{\ref{\Lew}} and conditions from modular 
covariance \q{\CaFus}. The latter require that the quantity 
$$
Z_{a b}(q) := \bdbadj\, \tqlc\,\bda = \sum_{i,j}\; \overline{B^i_b}\,
B^i_a\,S_{ij} \chi_j(q) = \sum_j\; n_{a b}^j \, \chi_j(q) 
\eq\Cardycon$$
can be interpreted as partition function of a CFT on a strip, with 
boundary conditions $a$ and $b$ imposed along the boundaries. In other 
words, the $n_{a b}^j$ must be non-negative integers. Above, we have 
introduced the conformal characters 
$\chi_j(q) := {\rm tr}_{{\cal H}_j} q\uplc$ of the irreducible modules 
$\h_j$, with $q= e^{2 \pi i \tau}$ and $\tilde q = e^{-2 \pi i/\tau}$, and 
the modular $S$-matrix from $\chi_i(\tilde q) = \sum_j S_{ij}\chi_j(q)$. 
\sn
For rational diagonal models, Cardy found a general solution to the 
constraints \Cardycon: Specializing to our case, the Cardy boundary 
states carry the same labels $a\in \ik$ from the Kac table as the 
Ishibashi states, and the coefficients are 
$$ 
B_a^i = {S_{a i}\over \sqrt{S_{0 i}} }
\eq\Cardybdst$$
($0 = (1,1)$ denotes the vacuum representation) with the minimal model 
$S$-matrix
$$
S_{(r,s)(r',s')}\ =\ {\textstyle 
\sqrt{{8\over m(m+1)}}\ (-1)^{1+rs'+sr'}\ 
\sin{(\pi{m+1\over m}rr')}\;\sin{(\pi{m\over m+1}ss')}  }\ \ .
\eq\Smat$$
The partition functions calculated with Cardy's boundary states have 
the simple form  
$Z_{a b}(q) = \sum_j\, N_{a b}^j\, \chi_j(q)$,  
where the minimal model fusion rules $N_{ij}^k$ can be read off from 
$$
\phi_{(r,s)} \times \phi_{(r',s')} = 
\sum_{{r''=|r-r'|+1 \atop r+r'+r''\ {\rm odd}}\ \ }^{r_{\rm max}} 
\sum_{{s''=|s-s'|+1 \atop s+s'+s''\ {\rm odd}}}^{s_{\rm max}} \ 
\phi_{(r'',s'')}
\eq\fusrules$$
with summations running up to $r_{\rm max}={\rm min}(r+r'-1,\;2m-r-r'-1)$ 
resp.\ $s_{\rm max}={\rm min}(s+s'-1,\;2m-s-s'+1)$. 
\sn
The Cardy boundary states $\bda$ form a complete set in the sense 
\q{\ref{\PSS}} that the matrix of coefficients $B_a^i$ is invertible; 
see also \q{\ref{\BPPZ}}. This implies that any other boundary 
state $\bdA$ is a (complex) linear combination of Cardy states. If 
we impose the natural compatibility condition that $Z_{A 0}(q)$, where 
$|\!|0\rangle\!\rangle$ is the Cardy state associated with the 
vacuum representation, is a partition function, then $\bdA$ is in fact 
an element of the $\Z_+$-lattice generated by the Cardy boundary states, 
$$
\bdA = \sum_{a\in\ik}\; n_A^a\,\bda\qquad{\rm with}\ \ n_A^a \in \Z_+\ \ .
\eq\superpos$$ 
\sn
The partition functions $Z_{a a}(q)$ describe the spectrum of boundary 
fields -- of field operators which can be inserted (only) along world-sheet 
boundaries where the boundary condition $a$ is imposed (in distinction to 
the bulk fields $\varphi_i(\zeta,\bzeta)$  which live in the interior 
of the disk). The boundary operators are in one-to-one correspondence 
to the space of states of the boundary CFT, for Cardy type boundary 
conditions given by 
$$
\h_{a a} = \bigoplus_{j\in\ik} \; \h_j^{\ \oplus\,N^j_{a a}}\ \ .
\eq\statesp$$
The $\h_j$ are irreducible representations of the Virasoro algebra,  
again with highest weights from the Kac table \Kact. Note that 
Cardy boundary conditions are distinguished among the  $\bdA$ 
by the property that $\h_{a a}$ contains the vacuum precisely 
once. 
\hbn
Partition functions $Z_{a b}(q)$ with $a\neq b$ are associated to 
spaces of so-called boundary condition changing operators, which induce 
a jump in the boundary condition. (In string theory, they correspond to 
excitations of open strings stretched between two different D-branes.) 
\sn
The decomposition \statesp\ in particular allows us to determine the set 
of relevant boundary operators: We have $h_m(r,s) < 1$ iff $s=r,r\pm1,r+2$, 
and the least relevant boundary field $\psi_{(1,3)}$ has conformal dimension 
$$
h_m{(1,3)} = {m-1\over m+1} \ \ .
$$
It appears in the state space of each Cardy boundary condition with the 
exception of the series $\bda= |\!|(r,1)\rangle\!\rangle$, for which 
$\h_{a a}$ contains no (non-trivial) relevant boundary fields at all. 
\sn
A conformal field theory is ``solved'' once all correlation functions 
are known. The $B_a^i$ determine \q{\ref{\CaL}} the one-point functions 
of primary bulk fields $\varphi_i(\zeta,\bzeta)$ in the presence of the 
boundary condition $a$, 
$$
\langle\,\varphi_i(\zeta,\bzeta)\,\rangle_a = 
{B_a^i \over |\,1-\zeta\bzeta\,|^{2h_i}}\ \ ,
$$
and with the help of the structure constants in the bulk operator product 
expansion and of the conformal Ward identities, one can in principle 
compute arbitrary correlators which involve only bulk fields. 
Correlators of (bulk and) boundary fields can be treated in the same 
way once the structure constants in the bulk-boundary OPE \q{\CaL} and 
in the OPE of boundary fields {\q{\Lew}} 
$$
\psi_k^{a b}(e^{i\theta_1}) \psi_l^{c d}(e^{i\theta_2}) \ \approx\ 
\sum_m\ \delta^{b,c}\, C^{a b d}_{k l m}\; 
(\theta_1-\theta_2)^{h_m-h_k-h_l}\;
\psi_m^{a d}(e^{i\theta_2}) 
\eq\bdyOPE$$
(for $\theta_1 \approx \theta_2$ and $0 \leq \theta_2 < \theta_1 < 2\pi$) 
are known. All these structure constants are subject to 
non-linear equations arising from duality or sewing constraints \q{\Lew}. 
For Virasoro minimal models, solutions for the constants in eq.\ \bdyOPE\ 
were constructed by Runkel \q{\ref{\Run}}:
$$
C_{k l m}^{a b c}= \Fus{b}{m}{a}{c}{k}{l}
\eq\bdyOPEconst$$
where {\sf F} is the fusing matrix of the (chiral) Virasoro model; 
explicit expressions for special entries are given in the Appendix, 
see \q{\Run} for the remaining cases. That the boundary OPE is related 
to the fusing matrix is a more general phenomenon, as was shown 
recently in \q{\ref{\FFFS}}; see also \q{\BPPZ} for earlier 
investigations. 
\sn
We have collected defining data of Virasoro minimal models on the 
unit disk. For what follows, the most important pieces of 
information are the one-point functions of the identity operator for 
Cardy boundary conditions $a=(r,s)$, 
$$
g_a \equiv B_a^0  = \bigl({\textstyle {8\over m(m+1)}}\bigr)^{1\over4}\
{ \sin{\pi r\over m} \sin{\pi s\over m+1}  
   \over \bigl(\,\sin{\pi\over m} \sin{\pi\over m+1}\,\bigr)^{1\over2}}\ \ ,
\eq\gfact
$$
the boundary OPE structure constants $C_{k k m}^{a a a}$ for the 
special case of a constant boundary condition, and the spectra 
of boundary (condition changing) operators contained in $\h_{a b}$. 
\bn\mn
{\bf 3. Relevant boundary perturbations}
\mn
Given a boundary CFT with correlation functions $\langle\,\cdot\,\rangle_a$, 
where $a$ denotes the boundary condition along the unit circle, we can 
select a boundary field $\psi$ and try to define a new set of 
correlators by the (formal) expression 
$$\eqalign{
 \langle \, \varphi_{i_1}&(\zeta_1,\bar \zeta_1) \cdots  
    \varphi_{i_N}(\zeta_N, \bzeta_N)\,\rangle_{a;\ \lambda\psi} 
\cr
  & \qquad\qquad =  \langle\, \varphi_{i_1}(\zeta_1,\bzeta_1) \cdots 
    \varphi_{i_N}(\zeta_N, \bzeta_N)\; P \exp \bigl\lbrace\, \lambda\, 
   \varepsilon^{h_\psi-1} S_\psi\,\bigr\rbrace  \,\rangle_{a} \ \ .
\cr}\eq\defcorr$$
The symbol $P$ denotes path ordering of the exponential of the 
perturbation 
$$
 S_\psi  :=\ \int_{\partial\Sigma} \psi(s) \, {d s} 
$$
inserted along the boundary of the world-sheet $\Sigma$ (here:\ a disk
of radius $L$). Path ordering takes care of the fact that boundary 
fields in general are not local wrt.\ each other or wrt.\ themselves 
-- cf.\ the specific ordering of world-sheet arguments and the 
appearance of the $\delta$-symbol for boundary condition changing 
operators in the expansion \bdyOPE. 
\hbn
The higher order terms of the exponential series require UV regularization 
e.g.\ by restricting all integrals to the region $|x_i-x_j| > \varepsilon$, 
where $\varepsilon$ is some UV cutoff. We introduce a dimensionless 
renormalization group parameter $l$ that regulates the scale of the 
theory, ${L\over\varepsilon}(l) = e^{-l}$. We will keep the disk 
circumference $2\pi L$ fixed so that we need not deal with IR
divergences; they could also be taken care of by introducing a finite 
temperature, see e.g.\ \q{\AfLtwo}.) 
\sn
The cutoff $\varepsilon$ was already used above to render the real 
parameter $\lambda$ in \defcorr\ dimensionless. $\lambda$ is the strength 
of the perturbation in direction of $\psi$. If the conformal dimension 
of $\psi$ is 1, i.e.\ if $\psi$ is marginal, then the new set of 
correlation functions may define a new boundary conformal field 
theory. A criterion which ensures that such a $\psi$ yields a truly 
marginal deformation to all orders in $\lambda$ was established 
in \q{\RSmod}, where also various general results on the 
structure of the new CFT were obtained, providing some insight into 
the moduli space of conformal boundary conditions (or of D-branes). 
\hbn
Here, we are interested in deformations with $h_\psi < 1$. Such 
relevant perturbations introduce dimensionful quantities into the 
theory and therefore break conformal invariance. The latter is restored 
only at fixed points of the renormalization group flow triggered by 
$\psi$, and we reach a new conformal boundary condition for the 
same bulk CFT that we started from. (Perturbations by 
boundary operators do not affect the local properties in the 
interior of the disk, thus the ``parent'' bulk theory 
is not changed.) 
\sn
Several approaches can be applied to exhibit the properties of the RG 
flow and of the new fixed points, e.g.\ TBA and TSCA methods. In this 
paper, we use (zero temperature) conformal perturbation theory (see 
\q{\ref{\CaLH}} for a nice introduction to the bulk case), which is 
suitable to study a small neighbourhood of the original theory $\lambda=0$. 
\sn
In this approach, new fixed points are determined from the zeroes 
of the beta function $\beta(\lambda) = d\lambda/d\ln(L/\varepsilon) = 
d\lambda/d l$, which in turn 
is computed perturbatively (in $\lambda$) from the OPE of the field 
$\psi$, see e.g.\ \q{\AfLone,\AfLtwo,\CaLH}, 
$$
\beta(\lambda) = (1-h_\psi) \lambda + C^{a a a}_{\psi\psi\psi} \lambda^2 
+ O(\lambda^3)\ \ .
\eq\betafct$$
Note that $\lambda$ denotes the renormalized coupling constant here. We 
have assumed that no other non-trivial relevant operators appear in the 
OPE of $\psi$ with itself, as is the case for the field $\psi=\psi_{(1,3)}$ 
we are going to study; otherwise, new counterterms arise which introduce 
additional coupling constants and render the analysis much more 
complicated; see \q{\CaLH}. 
\hbn
To this lowest non-trivial order, $\beta(\lambda^*) = 0$ yields
a fixed point at \q{\AfLone} 
$$
\lambda^* = - {y \over C^{a a a}_{\psi\psi\psi}} \quad\quad 
{\rm with}\ \ y := 1-h_\psi \ \ .
\eq\fixpt$$
This value can be inserted into the perturbative expansion of the
correlators \defcorr, and we can in particular determine the 
one-point function of the identity $g_a(\lambda)\equiv 
g_{a;\,\lambda\psi}$ at 
the new fixed point. In the work \q{\AfLone}, this quantity was 
interpreted as ground-state degeneracy of the boundary CFT with 
boundary condition $a$, and it was conjectured that $g_a(\lambda)$ 
decreases along the RG trajectory; see also  \q{\AfLtwo} for a 
perturbative proof of this ``g-theorem''. The formula \q{\AfLone}
$$
\ln\bigl(g_a(\lambda)\bigr) = \ln(g_a)
- \pi^2\,y\,C_{\psi\psi\one}^{a a a}\; \lambda^2
- {2\pi^2\over 3}\,C_{\psi\psi\psi}^{a a a}\,C_{\psi\psi\one}^{a a a}
\; \lambda^3 +O(\lambda^4)
\eq\goflam$$
is obtained from the general prescription \defcorr\ in a similar way 
as $\beta(\lambda)$; while only the OPE (more precisely the 
fusion channel yielding $\psi$ from $\psi\times\psi$) plays a role 
for the latter, we have to integrate two- and three-point functions 
of the perturbing field to arrive at $g_a(\lambda)$ and eq.\ \goflam. 
At the fixed point, the logarithm of the ground-state degeneracy is 
shifted from the original value by 
$$
\Delta\ln\bigl(g_a(\lambda^*)\bigr)= -{\pi^2\over 3}\;
{C_{\psi\psi\one}^{a a a}\over \bigl(\,C_{\psi\psi\psi}^{a a a}\,\bigr)^2 }
\;y^3\ \ .\eq\delgpert$$
Due to the specific coefficients in eqs.\ \fixpt\ and \goflam, this 
is exact up to $O(y^4)$-corrections -- as long as the coefficients 
of $\lambda^N$ in the terms we ignore have no higher order poles 
in $y$. These coefficients $I_N(y,\varepsilon)$ are the integrals 
of the $N$-point functions of the perturbing field $\psi$, and their 
singularities in $y$ have been studied for the $\varphi_{(1,3)}$-bulk 
deformation in \q{\ref{\LuCa}}. The arguments used there depend only
on the OPE of $\psi$ and can be applied to our situation, too. It 
follows that the $I_N(y,\varepsilon)$ are regular as $y\to 0$. 
\sn
Let us now specialize to Virasoro minimal models. In view of the remarks 
made in the previous section, we know that the boundary CFT at the fixed 
point \fixpt\ can again be described by a ``pure'' Cardy boundary 
state or by an integer linear combination \superpos\ thereof. Thus, 
the new ground-state degeneracy $g(\lambda^*)$ must be contained in 
(or be a sum of values in) the list \gfact, at least to accuracy 
$O(y^4)$. If we focus on perturbations by the least relevant boundary 
field $\psi=\psi_{(1,3)}$ in the boundary model $M_m$ with some 
boundary condition $\bda \neq |\!|(r,1)\rangle\!\rangle$, this 
precision becomes high for large values of $m$, since $y=2/(m+1)$. 
\sn
In order to determine the new boundary condition at the fixed point 
$\lambda^*$, we first expand the ratio of two ground-state degeneracies
$g_a$, $g_b$ from eq.\ \gfact\ in powers of $1/(m+1)$: 
$$\eqalign{
\ln\Bigl({g_{b}\over g_a}\Bigr)&= 
  \ln\Bigl({\beta_1\beta_2\over\alpha_1\alpha_2}\Bigr) 
 + {\pi^2\over 6}\,(\alpha_1^2+\alpha_2^2-\beta_1^2-\beta_2^2)\;
({m+1})^{-2} \phantom{xxxxxxxxxx}
\cr
&\phantom{xxxxxxxxxxxxx} +{\pi^2\over 3}\,(\alpha_1^2-\beta_1^2)\;
({m+1})^{-3} +O\bigl((m+1)^{-4}\bigr)\ \ , 
\cr}\eq\delgexp$$
where we have used the notation $a=(\alpha_1,\alpha_2)$, 
$b=(\beta_1,\beta_2)$, and where we assume that the 
labels for the boundary conditions are close to the origin 
of the conformal grid, i.e.\ $\alpha_i, \beta_i \ll m$. 
(Note that the next formula shows that if the $\alpha_i$ satisfy 
this restriction then so do the $\beta_i$.)
\hbn
If we plug in Runkel's OPE structure constants from the Appendix into 
the perturbative formula \delgpert, we obtain 
$$
\Delta\ln\bigl(g_a(\lambda^*)\bigr)= 
-{\pi^2\over 3}\;(\alpha_2^2 -1)\;({m+1})^{-3}
+ O\bigl((m+1)^{-4}\bigr)\ \ . 
\eq\delgins$$
Now we compare the expressions \delgexp\ and \delgins\ order by 
order in $1/(m+1)$ to determine the new boundary condition 
$b=(\beta_1,\beta_2)$; one finds the equations 
$$
\alpha_1\alpha_2=\beta_1\beta_2\ , \quad\ 
\alpha_1^2+\alpha_2^2=\beta_1^2+\beta_2^2\ , \quad\ 
\beta_1^2-\alpha_1^2=\alpha_2^2-1\ . 
\eq\conds$$
If the starting boundary condition $a=(\alpha_1,\alpha_2)$ 
satisfies $\alpha_1=1$, there is a (unique) solution over the 
positive integers,  
$$
\beta_1 = \alpha_2\ , \quad\ \beta_2=\alpha_1 \ , 
$$  
which corresponds to a flow from boundary condition $a=(1,r)$ to 
$b=(r,1)$. For specific small values of $m$, these RG trajectories 
have already been found by TBA methods \q{\TBAetc,\Chim}.
\sn
For $a\neq (1,r)$, one cannot find a Cardy boundary state $\bdb$ with 
$b\in \ik$ such that eqs.\ \conds\ are fulfilled. This means that we 
have to pass to superpositions of Cardy boundary states as in eq.\ 
\superpos, and to replace $g_b$ in \delgexp\ by a sum 
$g_{\rm sup}=g_{b_1} + \ldots + g_{b_N}$ with $g_{b_l}$ from eq.\ 
\gfact, each $b_l = (\beta^{l}_1,\beta^{l}_2)$ possibly 
occuring more than once. Then we get, after some algebra,  
$$\eqalign{
\ln\Bigl({g_{\rm sup}\over g_a}\Bigr)
&= \ln(\sigma)+
 {\pi^2\over 6}\;\Bigl(\,\alpha_1^2+\alpha_2^2 - \sum_l 
  {s_l\over\sigma}\,
  \bigl((\beta_1^{l})^2+(\beta_2^{l})^2\bigr)\,\Bigr)\;({m+1})^{-2} 
\phantom{xxxxxxxxx}\cr
&\quad\quad\quad\quad\quad\quad 
+{\pi^2\over 3}\; \Bigl(\,\alpha_1^2-\sum_l {s_l\over\sigma}\,
(\beta_1^{l})^2\Bigr)\;({m+1})^{-3}    +O\bigl((m+1)^{-4}\bigr)\ \ , 
\cr}\eq\delgsupexp$$
where we have introduced the abbreviations 
$s_l := {\beta_1^{l}\beta_2^{l}\over\alpha_1\alpha_2}$ and 
$\sigma:=\sum_l s_l$. Now we can again compare with the perturbative 
result \delgins\ and obtain the equations
$$
\sigma=1\ , \quad\
\sum_l\; s_l\Bigl((\beta_1^{l})^2+(\beta_2^{l})^2\Bigr)
         =\alpha_1^2+\alpha_2^2\ , 
\quad\
\sum_l\; s_l(\beta_2^l)^2= 1 \ \ .
\eq\mconds
$$
These enforce $\beta_2^l=1$ for all $l$, thus the RG flow triggered by 
the least relevant boundary field leads from a Cardy boundary condition 
$a$ to a superposition of $b_l= (\beta^l_1, 1)$. 
Depending on the actual values of $\alpha_1,\alpha_2$ and $m$, eqs.\ 
\mconds\ may admit several solutions for the labels $\beta^l_1$. 
There is, however, the distinguished generic solution 
$$
\beta^l_2 = 1\ ,\quad\ 
\beta_1^l = \alpha_1 + \alpha_2 + 1 - 2l\quad\ \ {\rm for}\quad 
 l = 1,\ldots,N := {\rm min}(\alpha_1,\alpha_2)\ \ .
\eq\gensupsol$$
A flow to a superposition of boundary conditions was already 
observed in \q{\Chim} for the special case $m=4$ and $a=(2,2)$. 
\sn
So far, all our calculations were based on the ground-state 
degeneracy. This leaves some ambiguities in the final boundary 
condition because of the symmetry $g_{(r,s)} = g_{(m-r,s)} = 
g_{(r,m+1-s)}$. We can, however, repeat the procedures from 
above for one-point functions of other bulk fields and compute 
e.g.\ the change of $\;\langle\varphi_{(2,2)}(\zeta,\bzeta)\rangle_{
a;\,\lambda\psi}$. This function picks up a sign under the 
transformations $a \equiv (r,s) \mapsto (m-r,s)$ or $a \mapsto 
(r,m+1-s)$. In fact, it is easy to see that the zeroth order in 
$\lambda$ already suffices to rule out $(m-\beta^l_1,\beta^l_2)$ 
and $(\beta^l_1,m+1-\beta^l_2)$ in favour of $(\beta^l_1,\beta^l_2)$ 
from eq.\ \gensupsol. With the help of higher order corrections, one can 
also exclude some of the non-generic solutions to eqs.\ \mconds. 
\bn\mn
{\bf 4. Concluding remarks}
\mn
Our perturbative analysis of relevant boundary deformations yields 
the following picture: The RG flow triggered by the least 
relevant boundary field $\psi_{(1,3)}$, starting from a Virasoro 
minimal model with Cardy boundary condition $a = (\alpha_1,\alpha_2)$, 
$\alpha_2>1$, has a non-trivial fixed point with new boundary condition 
given by a 
superposition of $N := {\rm min}(\alpha_1,\alpha_2)$ Cardy 
boundary states, namely 
\def\loarrr{\raise3.85pt\hbox{$\underline{\phantom{mxxxx}}$}
\!\!\!\longrightarrow}
\def\longarrr{\raise3.85pt\hbox{$\underline{\phantom{mxxxx}}$}
\!\!\!\longrightarrow}
\def\psiflow{\hbox{
            \raise6pt\hbox{$\psi_{(1,3)}$}  \hskip-35pt
            \raise-2pt\hbox{$\loarrr$}   }}
\def\minpsiflow{\hbox{
            \raise6pt\hbox{$-\psi_{(1,3)}$}  \hskip-39pt
            \raise-2pt\hbox{$\longarrr$}   }}

$$
a= (\alpha_1,\alpha_2) \quad\ \   \psiflow \quad 
a_{\lambda^*\psi_{(1,3)}} = \sum_{l=1}^N\  (\alpha_1+\alpha_2+1-2l,\;1)\ \ .
\eq\supflow$$
\sn
For the time being, this result remains slightly conjectural in that 
we cannot exclude all the non-generic solutions to eqs.\ \mconds\ by 
analytic means. In order to back up the RG flow pattern \supflow, one 
can resort to TCSA calculations which rest on an explicit 
diagonalization of the perturbed Hamiltonian in a finite-dimensional 
subspace of low-lying energy levels. This allows one to determine 
the spectrum $Z_{0,a_{\lambda^*\psi}}(q)$, from which the new 
boundary condition $a_{\lambda^*\psi}$ can be inferred by counting 
degeneracies of energy levels. This method is restricted 
to case-by-case studies, but in all examples tested so far the 
pattern \supflow\ was confirmed \q{\ref{\GWa}}, even for small 
$m$-values down to $m=3$. Likewise, the assumption $\alpha_i \ll m$, 
which we had to make for technical reasons, does not seem to play 
any role in the end. (When dropping this condition, the conformal 
grid symmetry makes the number $N$ of superimposed boundary conditions 
introduced in \gensupsol\ ambiguous: We should then pass to 
$(m-\alpha_1,m+1-\alpha_2)$ if one of the entries in this labeling 
is smaller than those in $(\alpha_1,\alpha_2)$.) 
\sn
TBA and TCSA investigations of $\psi_{(1,3)}$ boundary flows 
reveal the existence of a second fixed point $\lambda^*_-$ besides 
the perturbative value $\lambda^*_+ := \lambda^*$ from eq.\ \fixpt.  
The values $\lambda^*_\pm$ have different signs and correspond to 
moving away from the original theory in directions $\pm \psi_{(1,3)}$. 
\hbn
{}From the works \q{\TBAetc,\Chim} and from a large number of TCSA 
calculations \q{\GWa}, one is led to consider the following pattern: 
For large enough $\alpha_i$, the boundary condition $a_{\lambda^*_-\psi}$ 
reached by perturbing $a = (\alpha_1,\alpha_2)$ with $-\psi_{(1,3)}$ 
coincides with $a'_{\lambda^*_+\psi}$ reached by perturbing 
$a' := (\alpha_1,\alpha_2-1)$ with $+\psi_{(1,3)}$ as above, i.e.\ \q{\GWa} 
$$
a= (\alpha_1,\alpha_2) \quad\ \   \minpsiflow \quad 
a_{\lambda^*_-\psi_{(1,3)}} =
\sum_{l=1}^N\  (\alpha_1+\alpha_2-2l,\;1) 
\eq\minsupflow$$
with $N:= \min(\alpha_1,\alpha_2-1)$. In hindsight, it is clear that our 
perturbative analysis could not uncover this second class of 
fixed points, since $\Delta\ln(g(\lambda^*_-))$ does not become 
small for large $m$. 
\sn
Remarkably, in all these cases, perturbation with the least relevant 
boundary field already leads to a ``stable'' boundary condition: 
The spectrum at the new fixed point follows from eqs.\ \Cardycon\ 
and \superpos, 
$$ 
Z_{\rm sup}(q) = \sum_{l,l'=1}^N \; n_{\rm sup}^{b_l}\,n_{\rm sup}^{b_{l'}}
\; Z_{b_l b_{l'}}(q)\ \ ,
\eq\nfppartf$$
and using formula \statesp\ together with the fusion rules \fusrules, 
we see that $\beta^l_2=1$ implies that no relevant boundary (condition 
changing) operators besides the identity are left in the spectrum. 
This phenomenon is to be contrasted to the chain of least relevant 
bulk flows considered in \q{\ZaPert}. Formula \nfppartf\ also shows
that there are $N$ boundary operators of dimension zero contained 
in the spectrum of the superposition. 
\sn
The Virasoro minimal models on the disk studied here are intimately 
connected to RSOS models on a cylinder, whose ``spin variables'' take 
values in an $A_m$ graph. Here, boundary conditions appear as additional 
restrictions on the spin variables in the layers near the cylinder ends 
\q{\ref{\BaS},\CaFus,\BPPZ}. The pattern \supflow\  suggests that, when 
a relevant perturbation with $\psi_{(1,3)}$ is turned on 
along the boundaries of the cylinder, the lattice model ``breaks up'' 
into $N^2$ independent sublattices, with boundary conditions 
$b_l,\,b_{l'}$ as in \gensupsol. (The lattice interpretation of 
the Cardy boundary condition $(\beta^l_1,1)$ is that the spins 
in the two outer layers are fixed to $\beta_1^l$ and $\beta_1^l+1$.) 
The partition function of such a subsystem is then given by the term 
$Z_{b_l b_{l'}}(q)$ from eq.\ \nfppartf. As we mentioned in the 
introduction, Affleck found that this ``splitting into subsystems'' 
signals the presence of ordered phases along the boundary \q{\Aff}. 
The meaning of the associated first order phase transitions within 
brane physics remains to be understood better. 
\sn
One should also try to generalize the investigations of Section 3 
to minimal models of the $N=2$ super Virasoro algebra. These occur 
as building blocks of phenomenologically interesting string backgrounds, 
namely of Gepner models. Boundary conditions in these superconformal 
theories \q{\ref{\RSGep}} are  closely related to 
D-branes in superstring compactifications on Calabi-Yau manifolds 
\q{\ref{\CYGep}}. Statements on relevant boundary flows 
in such models should directly translate into statements on 
stability and bound state formation of the corresponding branes, 
extending ideas in \q{\Sen,\GNS} and the SU(2)-findings of \q{\ARS}.
\bn\sn
{\bf Acknowledgments}\qquad We are grateful to W.\ Nahm, P.\ Pearce, 
V.\ Rittenberg, G.\ von Gehlen and J.-B.\ Zuber 
for useful discussions. We are particularly indebted to G.\ Watts 
for valuable suggestions and for making his unpublished results 
on the TCSA analysis of relevant boundary flows available to us. 
D.R.\ thanks the AEI Potsdam for hospitality. A.R.\ thanks King's 
College, London, for hospitality during a visit supported by the 
TMR European Superstring Theory Network. 
\bn\mn
{\bf Appendix}
\mn
Here we give explicit expressions for specific structure constants 
of the boundary OPE of Virasoro minimal models. According to Runkel's 
work \q{\Run}, they follow from the fusing 
matrices which describe a change of basis in the space of conformal 
blocks. We write $a = (\alpha_1,\alpha_2)$ for the boundary condition 
as before and introduce $\delta_a:=\alpha_2-\alpha_1$, along with 
some further abbreviations: 
$$\eqalign{ 
A(\delta_a, \alpha_2)\ &:= \  {\ts
       \Gamma(\delta_a-{\alpha_2\over m+1})\;
       \Gamma(\delta_a-{\alpha_2-1\over m+1})\;
       \Gamma(1-\delta_a+{\alpha_2\over m+1})  }\ \ ,
\cr
B(\delta_a, \alpha_2)\ &:=\ {\ts
       \Gamma^2(1+\delta_a-{\alpha_2+1\over m+1})\;
       \Gamma(1-\delta_a+{\alpha_2-1\over m+1}) }\ \ ,
\cr
C(\delta_a, \alpha_2)\ &:=\ {\ts
        \Gamma^2(1+\delta_a-{\alpha_2+1\over m+1})\;
        \Gamma(-\delta_a+{\alpha_2+1\over m+1}) }\ \ ,
\cr
D(\delta_a, \alpha_2)\ &:=\  {\ts
           \Gamma(1+\delta_a+{\alpha_2-1\over m+1})\;
           \Gamma(-1+\delta_a-{\alpha_2-2\over m+1})\;
           \Gamma(2-\delta_a+{\alpha_2-2\over m+1})  }\ \ .
\cr}$$
Using the recursive formula for the {\sf F}-entries given in 
\q{\Run}, the structure constants needed for the investigation 
of relevant perturbations with $\psi = \psi_{(1,3)}$ can be written 
as follows: 
$$\eqalign{
C_{\psi\psi\one}^{a a a} &=  
   {\ts {\Gamma(2-{2\over m+1})\Gamma(2-{3\over m+1})\Gamma(1-{2 \over m+1})
                    \over
         \Gamma(1-{1\over m+1})} }
\;\Bigl(\, {A(\delta_a, \alpha_2) \over B(\delta_a, \alpha_2)} 
      + {A(- \delta_a, -\alpha_2) \over B(-\delta_a, -\alpha_2)}\,\Bigr) \ \ ,
\cr
C_{\psi\psi\psi}^{a a a} &\vphantom{\sum^M}= 
  {\ts {\Gamma^2(-1+{2\over m+1})\Gamma(-1+{3\over m+1})
        \Gamma(2-{2 \over m+1})\Gamma(2-{3\over m+1})
                    \over 
 \Gamma(-2+{4\over m+1})\Gamma^2({1\over m+1})\Gamma^2(1-{1\over m+1})} }
\;\Bigl(\, {A(\delta_a, \alpha_2) \over C(\delta_a, \alpha_2)} 
          + {A(- \delta_a, -\alpha_2) \over C(-\delta_a, -\alpha_2)}\,\Bigr)
\cr
&\qquad\quad  + \ 
   {\ts{ \Gamma(-2+{3\over m+1})\Gamma(3-{4\over m+1})\Gamma(3-{3\over m+1})
                  \over 
         \Gamma(1-{1\over m+1})\Gamma({1\over m+1})\Gamma(2-{2\over m+1})}  }
\;\Bigl(\, {A(\delta_a, \alpha_2) \over D(\delta_a, \alpha_2)} 
          + {A(- \delta_a, -\alpha_2) \over D(-\delta_a, -\alpha_2)}\,\Bigr)
\ \ .
\cr}$$
Now, we can expand the ratio $\;{C_{\psi\psi\one}^{a a a}/\bigl(\,
C_{\psi\psi\psi}^{a a a}\,\bigr)^2}\;$ in $1/(m+1)$, with the rather 
simple result
$$
{C_{\psi\psi\one}^{a a a}\over \bigl(\,C_{\psi\psi\psi}^{a a a}\,\bigr)^2 }=
{1\over 8}\;(\alpha_2^2-1) + O((m+1)^{-1})\ \ .
$$
\bn\sn
\eject\noindent
{\bf References}
\sn
\def\bf{\bfk}\def\sl{\itk}
\noindent
{\baselineskip=10pt {\klein 
\def\q#1{\vskip2.5pt\noindent\item{{\lb\klein #1\rb}}}
\q{\AfLtwo} I.\ Affleck, A.W.W.\ Ludwig, {\sl Exact conformal-field-theory 
  results on the multichannel Kondo effect: Single-fermion Green's function, 
  self-energy, and resistivity}, Phys.\ Rev.\ B{\bf48} (1993) 7297
\q{\FLuS} P.\ Fendley, A.W.W.\ Ludwig, H.\ Saleur, {\sl Exact conductance 
  through point contacts in the ${\scst \nu =1/3}$ fractional quantum 
  Hall effect}, Phys.\ Rev.\ Lett.\ {\bf74} (1995) 3005, cond-mat/9408068; \ \ 
 {\sl Exact non-equilibrium transport through point contacts in quantum 
  wires and fractional quantum Hall devices}, Phys.\ Rev.\ B{\bf52} (1995) 
  8934, cond-mat/9503172;\ \ {\sl Exact non-equilibrium DC shot noise in 
  Luttinger liquids and fractional quantum Hall devices}, Phys.\ Rev.\ 
  Lett.\ {\bf75} (1995) 2196, cond-mat/9505031
\q{\GNS} E.\ Gava, K.S.\ Narain, M.H.\ Sarmadi, {\sl On the bound states 
  of p- and (p+2)-branes}, Nucl.\ Phys.\ B{\bf504} (1997) 214, 
  hep-th/9704006
\q{\Sen} A.\ Sen, {\sl SO(32) spinors of type I and other solitons on 
   brane-antibrane pair}, J.\ High Energy Phys.\ 09 (1998) 023, 
  hep-th/9808141; \quad {\sl Descent relations among bosonic D-branes}, 
  Int.\ J.\ Mod.\ Phys. A{\bf14} (1999) 4061, 
  hep-th/9902105;\quad {\sl Non-BPS states and branes in string theory}, 
  hep-th/9904207
\q{\SZB} A.\ Sen, B.\ Zwiebach, {\sl Tachyon condensation in string field 
  theory}, hep-th/9912249
\noindent\item{} N.\ Berkovits, A.\ Sen, B.\ Zwiebach, {\sl Tachyon 
  condensation in superstring field theory}, hep-th/0002211
\noindent\item{}  J.A.\ Harvey, P.\ Kraus, {\sl D-Branes as unstable 
  lumps in bosonic open string field theory}, hep-th/0002117
\q{\HKM} J.A.\ Harvey, D.\ Kutasov, E.J.\ Martinec, {\sl On the 
  relevance of tachyons}, hep-th/0003101
\q{\TBAetc} P.\ Fendley, H.\ Saleur, N.P.\ Warner, {\sl 
 Exact solution of a massless scalar field with a relevant boundary
 interaction}, Nucl.\ Phys.\ B{\bf430} (1994) 577, hep-th/9406125%
\noindent\item{} P.\ Dorey, A.\ Pocklington, R.\ Tateo, G.\ Watts, 
  {\sl TBA and TCSA with boundaries and excited states}, 
  Nucl.\ Phys.\ B{\bf525} (1998) 641, hep-th/9712197
\noindent\item{} F.\ Lesage, H.\ Saleur, P.\ Simonetti, {\sl Boundary 
  flows in minimal models}, Phys.\ Lett.\ B{\bf427} (1998) 85, hep-th/9802061
\noindent\item{} C.\ Ahn, C.\ Rim, {\sl Boundary flows in general coset 
  theories}, J.\ Phys.\ A{\bf32} (1999) 2509, hep-th/9805101%
\noindent\item{} P.\ Dorey, I.\ Runkel, R.\ Tateo, G.\ Watts, {\sl g-function 
   flow in perturbed boundary conformal field theories}, hep-th/9909216
\q{\Chim} L.\ Chim, {\sl Boundary S-matrix for the tricritical 
  Ising model}, Int.\ J.\ Mod.\ Phys.\ A{\bf 11} (1996) 4491, hep-th/9510008
\q{\AfLone} I.\ Affleck, A.W.W.\ Ludwig, {\sl Universal noninteger 
  'groundstate  degeneracy' in critical quantum systems}, 
   Phys.\ Rev.\ Lett.\ {\bf67} (1991) 161
\q{\ZaPert} A.B.\ Zamolodchikov, {\sl Renormalization group and perturbation 
  theory about fixed points in two-dimensional field theory}, 
  Sov.\ J.\ Nucl.\ Phys.\ {\bf46} (1987) 1090
\q{\RSmod} A.\ Recknagel, V.\ Schomerus, {\sl Boundary deformation 
   theory and moduli spaces of D-branes}, Nucl.\ Phys.\ B{\bf 545} (1999) 
   233, 
   hep-th/9811237; \quad {\sl Moduli spaces of D-branes in 
  CFT-backgrounds}, hep-th/9903139
\q{\Aff} I.\ Affleck,  {\sl Edge critical behaviour of the
  2-dimensional tri-critical Ising model}, cond-mat/0005286
\q{\AOS} I.\ Affleck, M.\ Oshikawa, H.\ Saleur, {\sl Boundary critical 
   phenomena in the three-state Potts model}, J.\ Phys.\ A{\bf31}
  (1998) 5827, cond-mat/9804117 
\q{\ARS}  A.Yu.\ Alekseev, A.\ Recknagel, V.\ Schomerus, 
  {\sl Brane dynamics in background fluxes and non-commutative
   geometry}, J.\ High Energy Phys.\ 05 (2000) 010, hep-th/0003187
\q{\Ish} N.\ Ishibashi, {\sl The boundary and crosscap states
   in conformal field theories}, Mod.\ Phys.\ Lett.\ A{\bf4} (1989) 251 
\q{\CaFus} J.L.\ Cardy, {\sl Boundary conditions, fusion rules
 and the Verlinde formula}, Nucl.\ Phys.\ B{\bf324} (1989) 581 
\q{\Lew} D.C.\ Lewellen, {\sl Sewing constraints for conformal field
 theories on surfaces with boundaries}, Nucl.\ Phys.\ B{\bf372} (1992) 654
\q{\PSS}  G.\ Pradisi, A.\ Sagnotti, Y.S.\ Stanev, {\sl Completeness 
   conditions for boundary operators in 2d conformal field theory}, 
   Phys.\ Lett.\ B{\bf381} (1996) 97, hep-th/9603097
\q{\BPPZ} R.E.\ Behrend, P.A.\ Pearce, V.B.\ Petkova, J.-B.\ Zuber,  
  {\sl Boundary conditions in rational conformal field theories}, 
 hep-th/9908036%
\q{\CaL}  J.L.\ Cardy, D.C.\ Lewellen, {\sl Bulk and boundary operators
 in conformal field theory}, Phys.\ Lett.\ B{\bf259} (1991) 274
\q{\Run} I.\ Runkel, {\sl Boundary structure constants 
 for the A-series Virasoro minimal models}, Nucl.\ Phys.\ B{\bf549} (1999) 
  563, 
  hep-th/9811178; \quad {\sl Structure constants for the D-series Virasoro 
  minimal models}, hep-th/9908046
\q{\FFFS} G.\ Felder, J.\ Fr\"ohlich, J.\ Fuchs, C.\ Schweigert, 
    {\sl The geometry of WZW branes}, hep-th/9909030; \quad
  {\sl Conformal boundary conditions and three-dimensional topological 
  field theory},  Phys.\ Rev.\ Lett.\ {\bf84} (2000) 1659, 
  hep-th/9909140;\ \  {\sl Correlation functions and 
  boundary conditions in RCFT and three-dimensional topology}, 
  hep-th/9912239
\q{\CaLH} J.L.\  Cardy, {\sl Conformal invariance and statistical
 mechanics}, Lectures given at the Les Houches Summer School in 
 Theoretical Physics, 1988 
\q{\LuCa} A.W.W.\ Ludwig, J.L.\ Cardy, {\sl Perturbative evaluation 
  of the conformal anomaly at new critical points with applications 
  to random systems}, Nucl.\ Phys.\ B{\bf285} (1987) 687
\q{\GWa} G.\ Watts, unpublished results 
\q{\RSGep} A.\ Recknagel, V.\ Schomerus, {\sl D-branes in 
    Gepner models}, Nucl.\ Phys.\ B{\bf531} (1998) 185, hep-th/9712186 
\noindent\item{} S.\ Govindarajan, T.\ Jayaraman, T.\ Sarkar,
  {\sl Worldsheet approaches to D-branes on supersymmetric cycles}, 
  hep-th/9907131
\noindent\item{} M.\ Naka, M.\ Nozaki, {\sl Boundary states in Gepner 
  models}, hep-th/0001037 
\noindent\item{} I.\ Brunner, V.\ Schomerus, {\sl D-branes at singular 
  curves of Calabi-Yau compactifications}, hep-th/0001132
\q{\CYGep} H.\ Ooguri, Y.\ Oz, Z.\ Yin, {\sl D-branes on
  Calabi-Yau spaces and their mirrors}, Nucl.\ Phys.\ B{\bf477}
  (1996) 407, hep-th/9606112
\noindent\item{} M.\ Gutperle, Y.\ Satoh, {\sl D-branes in Gepner models 
  and supersymmetry}, Nucl.\ Phys.\  B{\bf543} (1999) 73, hep-th/9808080;
  \ \  {\sl D0-branes in Gepner models and \hbox{${\scst N=2}$} black 
  holes}, Nucl.\ Phys.\ B{\bf555} (1999) 477, hep-th/9902120
\noindent\item{} I.\ Brunner, M.R.\ Douglas, A.\ Lawrence, C.\ R\"omelsberger,   {\sl D-branes on the quintic}, hep-th/9906200
\noindent\item{} D.-E.\ Diaconescu, C.\ R\"omelsberger, {\sl D-branes and 
   bundles on elliptic fibrations}, hep-th/9910172
\noindent\item{} P.\ Kaste, W.\ Lerche, C.A.\ L\"utken and J.\ Walcher,
  {\sl D-branes on K3-fibrations}, hep-th/9912147
\noindent\item{} E.\ Scheidegger, {\sl D-branes on some one- and two-parameter 
  Calabi-Yau hypersurfaces}, hep-th/9912188
\q{\BaS} H.\ Saleur, M.\ Bauer, {\sl On some relations between local height 
  properties and conformal invariance}, Nucl.\ Phys.\ B{\bf320} (1989) 591%
\smallskip}}
\bye